\documentclass[aps,twocolumn,prd,showpacs,showkeys,preprintnumbers,superscriptaddress,bibnotes,floatfix,longbibliography,notitlepage,nofootinbib,groupedaddress,reprint]{revtex4-2}

\usepackage[english]{babel}
\usepackage{amsfonts}

\DeclareSymbolFontAlphabet{\mathbb}{AMSb}

\usepackage{enumitem}
\usepackage{braket}
\usepackage{amsmath,amssymb}
\usepackage{txfonts}
\usepackage{comment}
\usepackage{bm}
\usepackage{mathtools}
\usepackage{makecell}

\usepackage[usenames, dvipsnames, svgnames, table]{xcolor}
\usepackage[colorlinks=true, citecolor=Blue,
            linkcolor=BrickRed, urlcolor=ForestGreen]{hyperref}
\usepackage[capitalise]{cleveref}

\usepackage{graphicx}
\usepackage{xfrac}
\usepackage{ulem}
\usepackage{cancel}

\usepackage[
raggedright
]{sidecap}   
\sidecaptionvpos{figure}{c} 

\begin{document}

\title{Beam displacement tolerances on a segmented mirror for higher-order Hermite-Gauss modes}

\overfullrule 0pt 
\parskip 0pt
\hyphenpenalty 9999

\author{Liu Tao}
\email{liu.tao@ligo.org} 
\affiliation{University of Florida, 2001 Museum Road, Gainesville, Florida 32611, USA}

\author{Nina Brown}
\affiliation{University of Chicago, Chicago IL 60637}

\author{Paul Fulda}
\affiliation{University of Florida, 2001 Museum Road, Gainesville, Florida 32611, USA}

\begin{abstract}
Odd-indexed higher-order Hermite-Gauss (HG) modes are compatible with 4-quadrant segmented mirrors due to their intensity nulls along the principal axes, which guarantees minimum beam intensity illuminating the bond lines between the segments thus leading to low power loss. However, a misplaced HG beam can cause extra power loss due to the bright intensity spots probing the bond lines. This paper analytically and numerically studies the beam displacement tolerances on a segmented mirror for the $\mathrm{HG_{3,3}}$ mode. We conclude that for ``effective'' bond lines with 6\,$\mu$m width, and the $\mathrm{HG_{3,3}}$ beam size chosen to guarantee 1\,ppm clipping loss when centered, the beam can be rotated by roughly 1 degree or laterally displaced by 4\,\% of its beam size while keeping the total power on the bond lines under 1\,ppm. We also demonstrate that the constrained beam displacement parameter region that guarantees a given power loss limit, or the beam displacement tolerance, is inversely proportional to the bond line thickness. 
\end{abstract}


\maketitle

Thermal noise of the test masses is one of the limiting noise sources in advanced gravitational wave (GW) detectors~\cite{aLIGO, AdVirgo, PhysRevD.102.062003}. It is expected to remain a limiting noise source in future detectors~\cite{Adhikari_2020, ET}, despite radical changes to the design including cryogenic operations, new materials, and the use of longer laser wavelengths~\cite{ PhysRevLett.127.071101, Meylahn_2022, Penn:19, Steinlechner:22}. There has been a continued research interest within the gravitational wave community in the so-called ``flat beams'', in particular higher-order Hermite-Gaussian (HG) beams as the probe beam for the detectors~\cite{Liu2020, Ast_Stefan_2021, tao2021power, Jones_2020, Heinze_Joscha_2022}, in place of the currently-used fundamental Gaussian beam, for their thermal noise benefits. With flatter beam intensity distributions, higher order beams can better average over the random mirror surface fluctuations caused by thermal motions, thus leading to lower thermal noise~\cite{Mours_2006, Vinet_2010}.

\begin{figure}[htbp]
    \centering
    \includegraphics[width=0.9\linewidth]{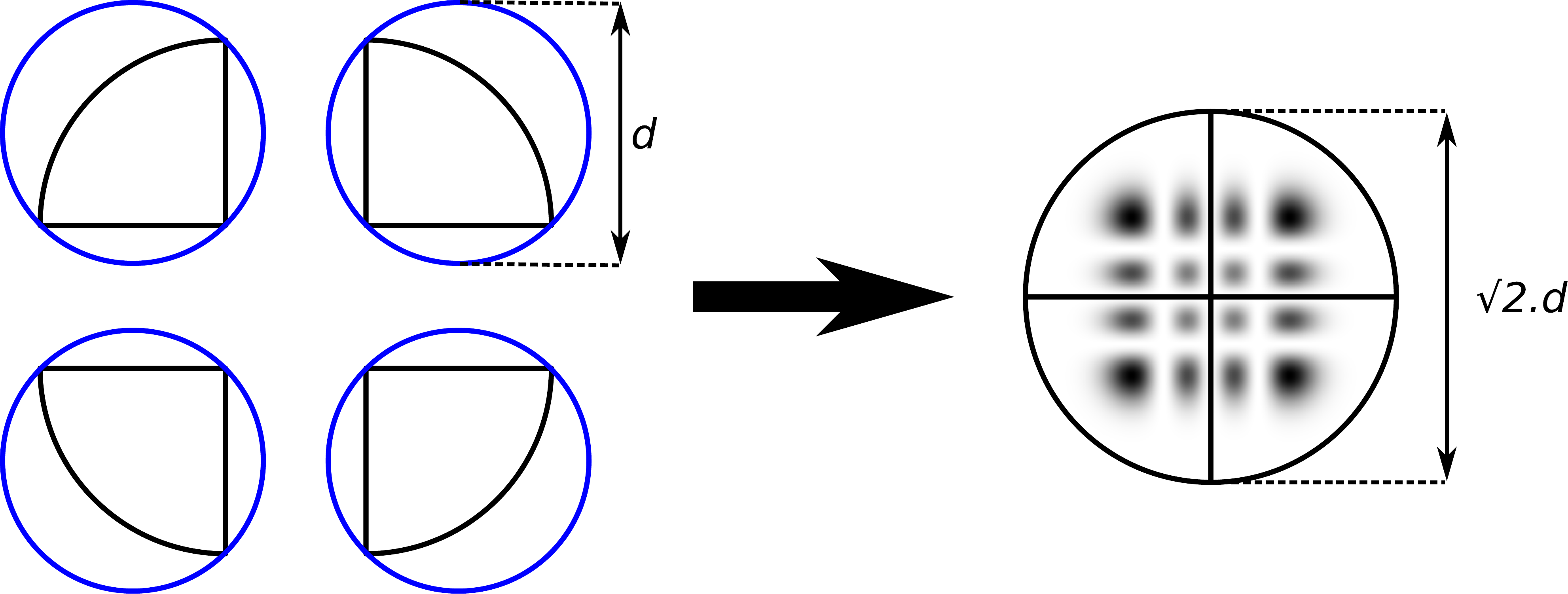}
    \caption{Illustration of an $\mathrm{HG_{3,3}}$ beam incident on a circular segmented mirror. The size of the segmented mirror is increased by a factor of $\sqrt{2}$, and the mass by a factor of $2\sqrt{2}$.}
    \label{fig: segmented}
\end{figure}

One other potential benefit of using higher-order HG modes is that one can take advantage of their intrinsic intensity pattern symmetries to use larger and more massive segmented mirrors, fabricated from smaller substrates that are limited by the ``boule size'' of high-purity silicon~\cite{Liu2020}. This optics segmentation design idea has also been studied by other people within the GW community in lowering the optics thermal noise, for instance, a ``piecewise'' coating design has been proposed by Kontos and Loglia~\cite{dcc-LIGO-P2300058} to satisfy high-quality coatings and relatively large optics diameters. In this paper, we consider one possible implementation of segmented optics, namely the 4-quadrant circular segmented mirror shown in Fig.~\ref{fig: segmented}. It is made by grouping four identical quadrants together, with each quadrant cut from a smaller ``boule''. The circular segmented mirror has its radius and thus the maximum allowable beam size assuming the same clipping loss requirement enlarged by a factor of $\sqrt{2}$, and its volume and therefore mass enlarged by a factor of $2\sqrt{2}$, assuming fixed dimensional ratios.

\begin{table}[htbp]
\centering
\caption{Coating and substrate Brownian thermal noise (CBTN and SBTN) and quantum radiation pressure noise (QRPN) power
spectral density (PSD) improvement~\cite{Vinet_2010, Bond2017} for the segmented mirror in Fig.~\ref{fig: segmented}.}
\begin{tabular}{c||c|c|c}
\hline \hline 
 Noise & CBTN & SBTN & QRPN \\
\hline \hline 
 PSD Scaling Relation & $1/w^2$ & $1/w$ & $1/M^2$ \\
 \hline \hline 
 Segmented Mirror Improvement & 2 & $\sqrt{2}$ & 8\\
\hline \hline
\end{tabular}
\label{tab:noise}
\end{table}

This allows a further reduction of thermal noises and quantum radiation pressure noise, as shown in Table~\ref{tab:noise}. For instance, the coating Brownian thermal noise power spectral density will be improved by a factor of 2 since it is inversely proportional to the square of the beam size at the mirror~\cite{Vinet_2010}. 

\begin{figure}[htbp]
    \centering
    \includegraphics[width=0.9\linewidth]{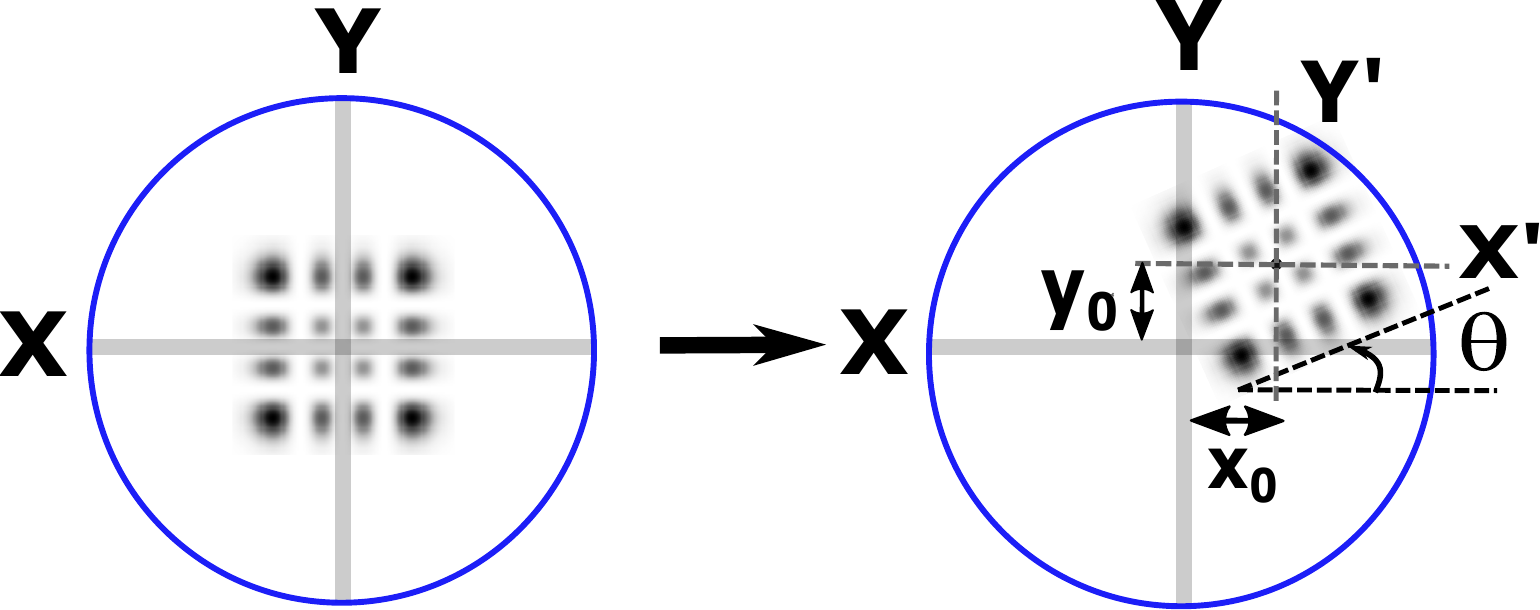}
    \caption{Illustration of an arbitrary displacement of an $\mathrm{HG}_{3,3}$ beam on a segmented mirror of the type shown in Fig.~\ref{fig: segmented}, characterized by a rotation of angle $\theta$ and lateral displacements of $x_{0}$ and $y_{0}$ along X and Y. The X and Y bond lines are shown as the grey strips along the principle axes.}
    \label{fig: displacement}
\end{figure}

The advantage of HG modes here comes from the fact that the intensity nulls can be aligned with the bond lines, thus avoiding interaction of light with the bonds and the associated scattered light penalties and thermal noise. If the beam is displaced on the segmented mirror, parts of the beam intensity will probe the bond lines though, as illustrated in Fig.~\ref{fig: displacement}. In this work we define the beam power on the bond lines as the \textit{power loss}, which we want to minimize or keep below some acceptable level.

This manuscript takes both analytical and numerical approaches to study the beam displacement tolerances for higher-order HG modes on the segmented mirror. We will start with an analytical approach that linearly expands the displaced beam in the segmented mirror basis, as denoted as ($X, Y$) in Fig.~\ref{fig: displacement}. The result provides valuable insights into the problem but is only valid for small deviations. When the deviation gets large, a numerical approach is required. We then adopt a complete numerical scheme that represents the displaced beams as discrete arrays. It gives quantitative arguments regarding what the typical ``effective'' bond line thickness is. Since extremely small quantities such as the power loss on the bond lines are involved, the grid convergence and the accuracy of our numerical scheme are discussed. We will look at how well the beam has to be centered and well aligned, in order to have at most 1 ppm of its power probing the bond lines, and how the beam displacement tolerances vary as we may have uncertainty in the ``effective'' bond line thickness due to the uncertainty in the beam propagation direction. The numerical results show good agreement with the
analytical results. We report our conclusions and discussions at the end.

For simplicity but without loss of generality, we consider a $\mathrm{HG}_{\mathrm{n},\mathrm{n}}$ mode at its waist, where n is odd in order to be compatible with segmented mirrors, see Fig.~\ref{fig: segmented}. We assume no clipping on the mirror for the analytical calculations and focus on the power loss on the bond lines in the presence of beam displacement.

As illustrated in Fig.~\ref{fig: displacement}, a generic beam displacement can be decomposed into a combination of lateral translations along $X$ and $Y$ of amount $x_{0}$ and $y_{0}$, and a counter-clockwise rotation of angle $\theta$ between the new axes and the original principle axes. Assuming reasonably small beam displacement ($x_{0}/w_{0}, y_{0}/w_{0}, \theta \ll 1$) as seen in real-life experiments, we expand the displaced beam up to the linear order in $x_{0}$, $y_{0}$, and $\theta$. We will see under such expansion, the lateral translations and the rotation do not cross-couple with each other and they contribute to the power loss on the bond lines independently. 

We first expand the rotated beam in the shifted basis centered at $(x_{0}, y_{0})$, denoted as ($X^{\prime}, Y^{\prime}$) in Fig.~\ref{fig: displacement}, and then expand each shifted beam component in the original basis ($X, Y$).

An $\mathrm{HG}_{\mathrm{n},\mathrm{n}}$ mode rotated by $\theta$ in ($X^{\prime}$, $Y^{\prime}$) basis looks like 
\begin{equation}
\begin{aligned}
\mathrm{HG_{n,n}}(x^{\prime},y^{\prime}) 
&\overset{\theta}{\Longrightarrow}
 \mathrm{HG_{n,n}} (\cos \theta x^{\prime} + \sin \theta y^{\prime}, -\sin \theta x^{\prime} + \cos \theta y^{\prime}) \\ &\approx \mathrm{HG_{n,n}} (x^{\prime} + \theta y^{\prime}, -\theta x^{\prime} + y^{\prime})
\end{aligned}
\label{equ:rotationtheta}
\end{equation}
where the small-angle approximation is used, and we denote the coordinates in the shifted ($X^{\prime}, Y^{\prime}$) basis as $x^{\prime}$ and $y^{\prime}$, scaled by the beam size.

Since HG modes are separable in the x and y axes, meaning $\mathrm{HG_{nm}}(x^{\prime}, y^{\prime})=\mathrm{U_{n}}(x^{\prime})\cdot \mathrm{U_{m}}(y^{\prime})$, we can treat them independently. Dealing with the x component first:
\begin{equation}
\begin{aligned}
    \mathrm{U_{n}}(x^{\prime}) &= \frac{1}{\sqrt{2^{n}n!}} H_{n}(\sqrt{2}x^{\prime})e^{-x^{\prime 2}} \\
    &\overset{\theta}{\Longrightarrow} \frac{1}{\sqrt{2^{n}n!}} H_{n}(\sqrt{2}(x^{\prime}+\theta y^{\prime}))e^{-(x^{\prime}+\theta y^{\prime})^2}\\
    &\approx \frac{1}{\sqrt{2^{n}n!}} \left(H_{n}(\sqrt{2} x^{\prime}) +2\sqrt{2}\theta y^{\prime} n H_{n-1}(\sqrt{2}x^{\prime})\right) e^{-x^{\prime 2}}\\
    &\times (1-2\theta x^{\prime}y^{\prime})
\end{aligned}
\label{equ:Ux}
\end{equation}
where we used the property $d H_{n}(x) /dx = 2nH_{n-1}(x)$ for Hermite polynomials~\cite{tao2021power}.

We have a similar result for the y component. In fact from Eq.~\ref{equ:rotationtheta} we can see that one can simply replace x with y and $\theta$ with $-\theta$ in Eq.~\ref{equ:Ux} to get the corresponding y component
\begin{equation}
\begin{aligned}
    \mathrm{U_{n}}(y^{\prime}) \overset{\theta}{\Longrightarrow} & \frac{1}{\sqrt{2^{n}n!}} \left(H_{n}(\sqrt{2} y^{\prime}) -2\sqrt{2}\theta x^{\prime} n H_{n-1}(\sqrt{2} y^{\prime})\right) e^{-y^{\prime 2}} \\
    &\times (1+2\theta x^{\prime}y^{\prime})
\end{aligned}
\label{equ:Uy}
\end{equation}

Combine Eqs.~\ref{equ:Ux} with~\ref{equ:Uy} we get the full decomposition of the rotated $\mathrm{HG}_{\mathrm{n},\mathrm{n}}$ mode in ($X^{\prime}, Y^{\prime}$) basis as
\begin{equation}
\begin{aligned}
    \mathrm{HG_{n,n}}(x^{\prime},y^{\prime})& =  \mathrm{U_{n}}(x^{\prime})  \mathrm{U_{n}}(y^{\prime}) \\
     &\approx \mathrm{HG_{n,n}} + \theta \sqrt{n(n+1)} \Big(\mathrm{HG_{n-1,n+1}} - \mathrm{HG_{n+1,n-1}}\Big) 
\label{equ: rotation}
\end{aligned}
\end{equation}
, all functions of $(x^{\prime}, y^{\prime})$. We also used Hermite polynomials property $2\sqrt{2}x H_{n} = H_{n+1} + 2nH_{n-1}$. As a result of rotation, the $\mathrm{HG_{n,n}}$ mode is scattered into $\mathrm{HG_{n-1,n+1}}$ and $\mathrm{HG_{n+1,n-1}}$ modes up to the linear order. This means that, if we start with an odd-indexed mode to have a dark stripe along the principle axes, such as the $\mathrm{HG}_{3,3}$ mode in Fig.~\ref{fig: segmented}, the rotation would scatter the original $\mathrm{HG}_{3,3}$ mode into even modes $\mathrm{HG}_{2,4}$ and $\mathrm{HG}_{4,2}$ modes. This causes extra beam intensity to probe the bond lines as they no longer have intensity nulls.

We now expand all the laterally shifted modes in Eq.~\ref{equ: rotation} in the original ($X, Y$) basis, where we have $x^{\prime} = x-x_{0}$ and $y^{\prime} = y-y_{0}$. From our previous work on the higher-order mode scattering due to lateral translations~\cite{tao2021power}, we have for the x component
\begin{equation}
\begin{aligned}
    \mathrm{U_{n}}(x) &\overset{x_{0}}{\Longrightarrow}
    \mathrm{U_{n}}(x-x_{0}) \\
    &= \mathrm{U_{n}}(x) + x_{0} \left( \sqrt{n+1}\mathrm{U_{n+1}}(x) - \sqrt{n}\mathrm{U_{n-1}}(x)\right)
\end{aligned}
\label{equ:Uxx}
\end{equation}
Similarly, for the y component, we have
\begin{equation}
\begin{aligned}
    \mathrm{U_{n}}(y) &\overset{x_{0}}{\Longrightarrow}
    \mathrm{U_{n}}(y-y_{0}) \\
    &= \mathrm{U_{n}}(y) + y_{0} \left( \sqrt{n+1}\mathrm{U_{n+1}}(y) - \sqrt{n}\mathrm{U_{n-1}}(y)\right)
\end{aligned}
\label{equ:Uyy}
\end{equation}
Combining Eqs.~\ref{equ:Uxx} and~\ref{equ:Uyy} we have
\begin{equation}
\begin{aligned}
    \mathrm{HG_{n,n}}(x,y) &\overset{x_{0}}{\Longrightarrow}
    \mathrm{HG_{n,n}} + x_{0} \left( \sqrt{n+1} \mathrm{HG_{n+1,n}} - \sqrt{n}\mathrm{HG_{n-1,n}}\right)  \\
    &+ y_{0} \left( \sqrt{n+1} \mathrm{HG_{n,n+1}} - \sqrt{n}\mathrm{HG_{n,n-1}}\right)
\end{aligned}
\label{equ: shift}
\end{equation}
up to the linear order.

Applying Eq.~\ref{equ: shift} to all shifted modes in Eq.~\ref{equ: rotation} we see that, up to the linear order, an arbitrarily displaced $\mathrm{HG_{n,n}}$ beam as shown in Fig.~\ref{fig: displacement} can be expanded in the original ($X, Y$) basis as
\begin{equation}
\begin{aligned}
\mathrm{HG_{n,n}}
&\overset{(x_{0},y_{0}),\theta}{\Longrightarrow}
    \mathrm{HG_{n,n}} + x_{0} \left( \sqrt{n+1} \mathrm{HG_{n+1,n}} - \sqrt{n}\mathrm{HG_{n-1,n}}\right)  \\
    &+ y_{0} \left( \sqrt{n+1} \mathrm{HG_{n,n+1}} - \sqrt{n}\mathrm{HG_{n,n-1}}\right) \\
    &+ \theta \sqrt{n(n+1)} \left( \mathrm{HG_{n-1,n+1}} - \mathrm{HG_{n+1,n-1}} \right)
\end{aligned}
\label{equ:final}
\end{equation}
We can see that up to the linear order, the lateral translations and the rotation contribute to mode scattering and thus the power loss on the bond lines independently. Since the lateral translation along the x and y have a similar contribution to the scattered mode contents due to symmetry, we will focus on the lateral translation along the x direction. With the scattered mode contents in Eq.~\ref{equ:final}, we can then square the result and integrate over the bond line regions to get the total power loss.

We now calculate the power loss on the bond lines numerically by representing
the displaced beam into discretized arrays. We consider a probe beam of $\mathrm{HG_{3,3}}$ mode with a total power of 1 W at its waist both angularly and laterally displaced on the circular segmented mirror. We consider an aLIGO-like test mass with a radius $R_{m}$ of 0.15\,m. The $\mathrm{HG_{3,3}}$ beam size is scaled to 0.0394\,m ($0.263 \cdot R_{m}$) to maintain 1\,ppm clipping loss, which is the beam power lost on the edge of the finite-aperture mirror.

We are interested in the ``effective'' bond line thickness considering the beam propagation directional uncertainties. In real-life experiments, the beam may not be exactly parallel to the bond plane. We expect this uncertainty in beam propagation direction $\theta_{beam}$ to be small, estimated as
\begin{equation}
\theta_{beam} = \Theta_{cav}
\end{equation}
where $\Theta_{cav}=\frac{\lambda}{\pi w_{0}}$ is the far-field beam divergence angle of aLIGO-like arm cavities. As the beam propagates through the thickness of the mirror substrate $t_{sub}$, 
it encounters the projection of the bond line onto its propagation axis: we refer to the projected width as the ``effective'' bond line width $d_{\mathrm{eff}}$~\footnote{Realistically $d_{\mathrm{eff}}$ would get reduced by the refraction index of silicon due to Snell's law. This is not considered here for an order-of-magnitude estimation.}
\begin{equation}
d_{\mathrm{eff}} = t_{sub} \cdot \theta_{beam} = \frac{t_{sub} \cdot \lambda}{\pi w_{0}} \approx 6 \mu m
\label{equ: bondlined}
\end{equation}
where $\rm{t}_{sub} \approx 20$\,cm and $w_{0} \approx 1.2$\,cm for aLIGO-like arm cavities.

\begin{figure}[htbp]
    \centering
    \includegraphics[width=0.9\linewidth]{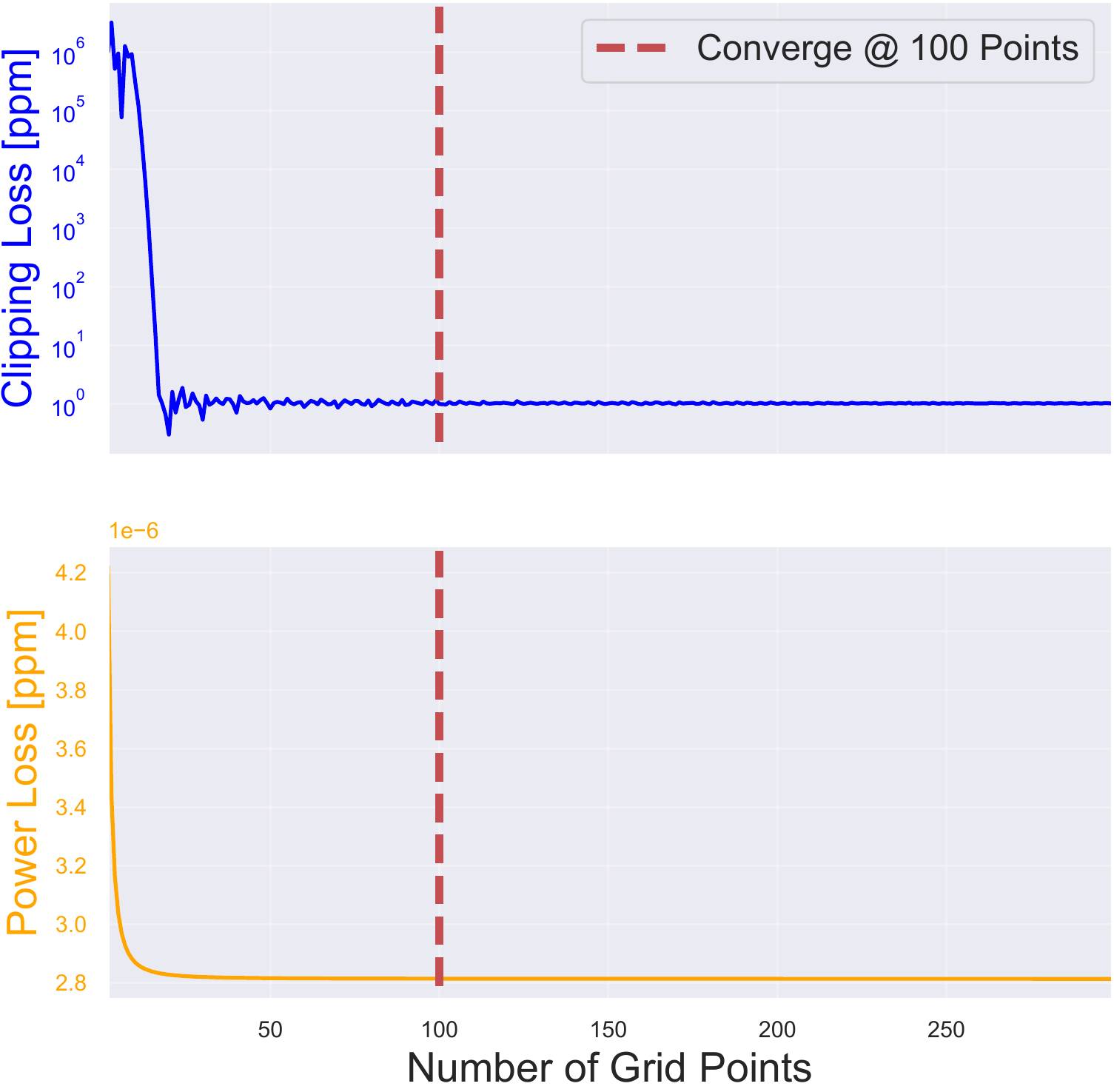}
    \caption{An image showing the grid convergence of our numerical scheme. Top: $\mathrm{HG_{3,3}}$ mode clipping loss on the entire mirror; Bottom: total power loss on the bond lines. In both cases, the power converges quickly as the number of grids increases, after around 100 cells.}
    \label{fig: gridconvergence_macro_micro}
\end{figure}

In our numerical approach, the displaced beams are discretized into arrays to calculate the power on 2D regions, the grid size has to be chosen carefully to guarantee numerical convergence and accuracy. To determine the optimal grid size, the clipping loss is first calculated as the number of grid points increases, as shown in the top panel of Fig.~\ref{fig: gridconvergence_macro_micro}. The clipping loss quickly converges to the desired 1 ppm after around 100 cells for the entire mirror. With the number of cells along the mirror diameter set to 100 cells, the power loss on the bond lines with a thickness of 6 $\mu m$ is calculated as the number of grid points along the bond line is increased, as shown in the bottom panel of Fig.~\ref{fig: gridconvergence_macro_micro}. The power loss also converges quickly after around 100 cells. We thus use 100 cells for both the mirror diameter and the bond line thickness in later calculations. We can also see that the power loss on the bond lines has small values of order $10^{-6}$ ppm with 6\,$\mu m$ thick bond lines, because of the intensity nulls of $\mathrm{HG_{3,3}}$, while the power loss for the $\mathrm{HG_{0,0}}$ is 170 ppm.

\begin{figure}[htbp]
    \centering
    \includegraphics[width=0.9\linewidth]{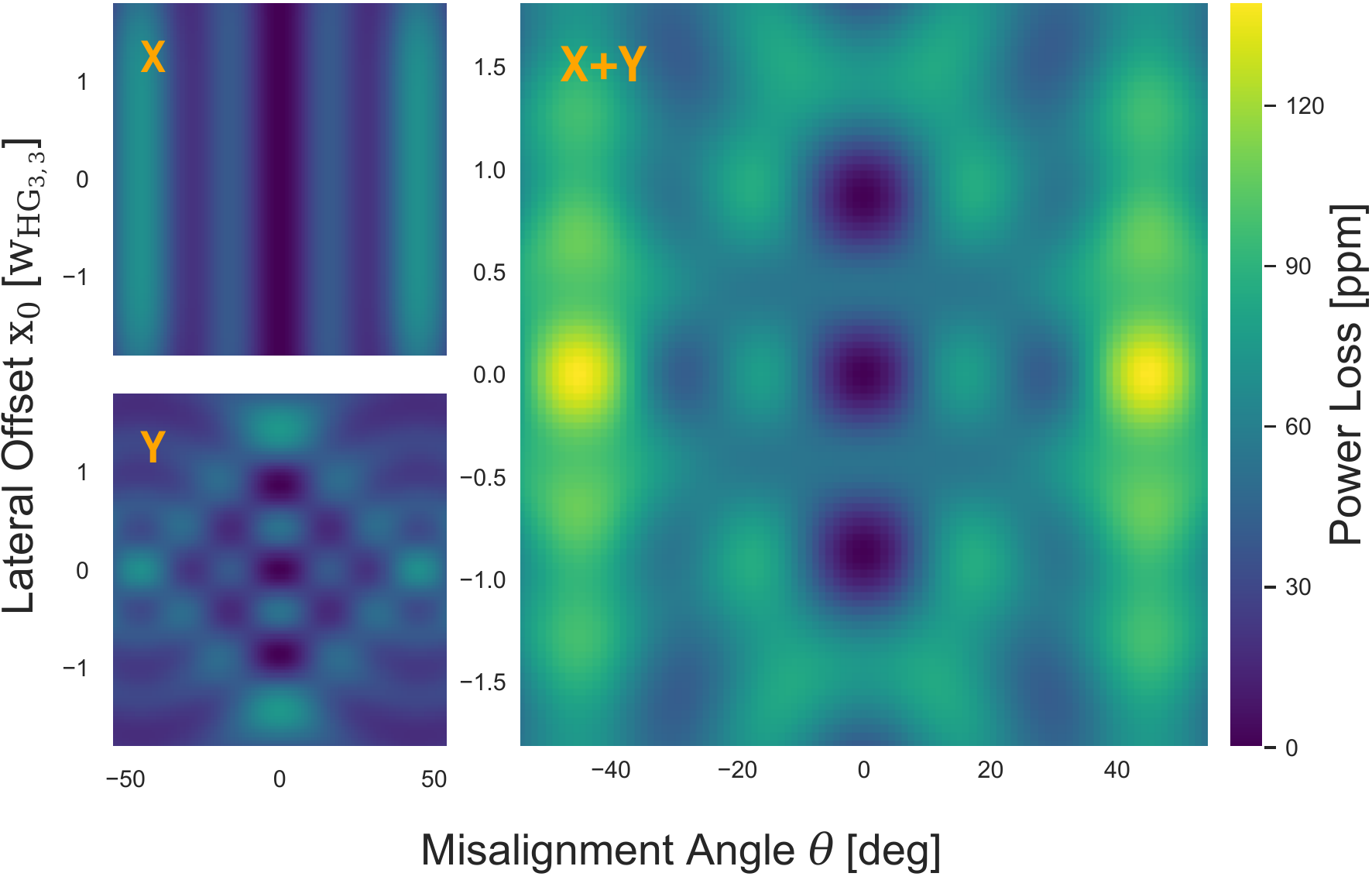}
    \caption{An image showing the power loss on the X and Y bond lines on the top and bottom left, and the total power loss on the right, due to lateral offset $x_{0}$ in the unit of the waist size $\mathrm{w_{HG_{3,3}}}$ and rotation $\theta$ in degrees.}
    \label{fig: PLXY_macro_micro}
\end{figure}

The probe beam $\mathrm{HG_{3,3}}$ mode is displaced both laterally along the x direction, in the unit of the beam size, and angularly in degrees. The total power loss on the X and Y bond lines is calculated by discretizing the displaced beam on the bond lines into matrices. The result is shown in Fig.~\ref{fig: PLXY_macro_micro}. We see as the beam is displaced with respect to the segmented mirror, different parts of the beam profile illuminate on the bond lines, which causes extra power losses. The power loss can go up to as high as 120 ppm for 6 $\mu m$ bond lines when the beam is rotated by 45 degrees. This corresponds to when the brightest spot of the $\mathrm{HG_{3,3}}$ beam at the corners hits the bond lines.

\begin{figure}[htbp]
    \centering
    \includegraphics[width=0.9\linewidth]{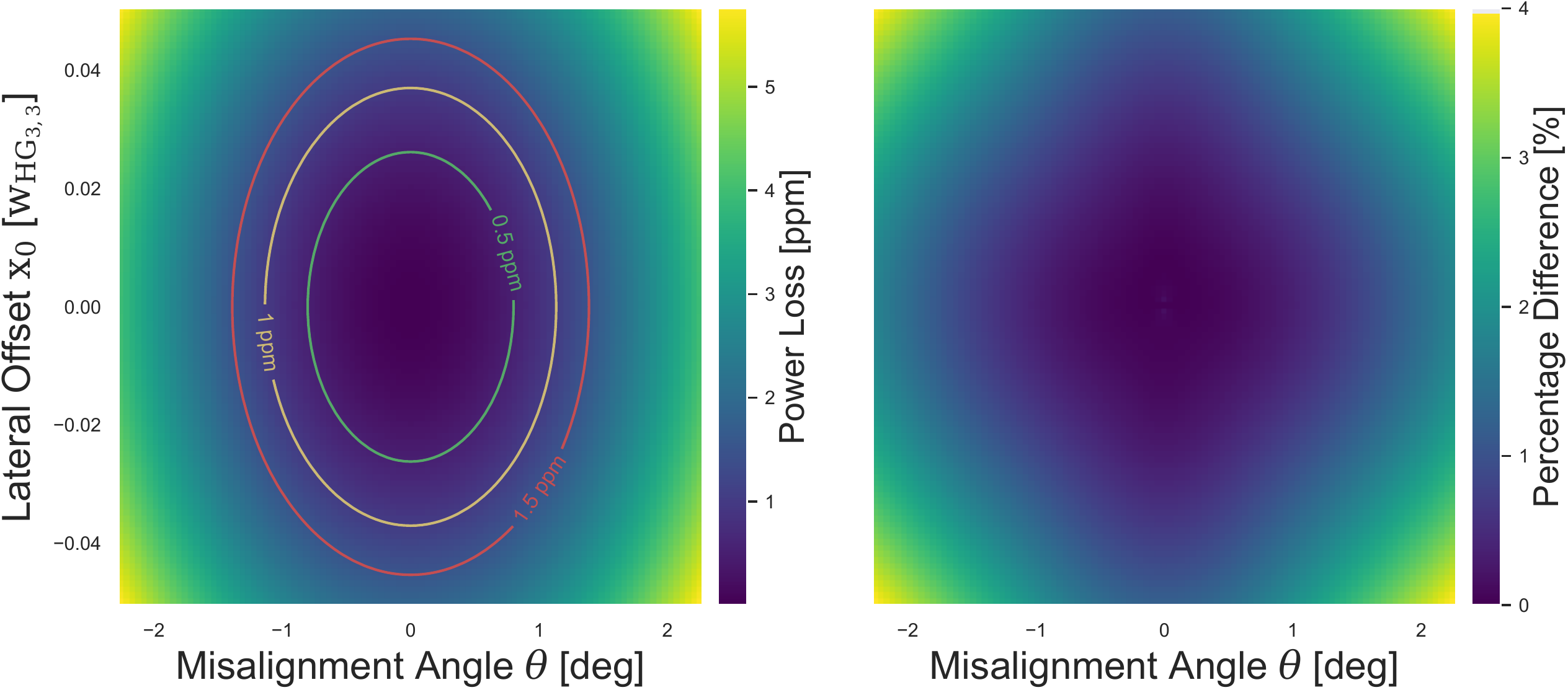}
    \caption{Left: the same total power loss as Fig.~\ref{fig: PLXY_macro_micro} but due to smaller beam displacements. Right: the percentage difference between the numerical result and the analytical result in Eq.~\ref{equ:final}. Three contours corresponding to total power loss of 0.5 ppm, 1 ppm, and 1.5 ppm are included on the left.}
    \label{fig: PLXY_micro_compare}
\end{figure}

In current GW detectors, however, the typical beam displacement would be much smaller. The total power loss due to small beam displacement is shown on the left in Fig.~\ref{fig: PLXY_micro_compare}. Three contours corresponding to power loss of 0.5 ppm, 1 ppm, and 1.5 ppm are included. We see that even if the beam is displaced by 1 degree angularly, or by 4\% of the beam size laterally, the total power loss on the 6 $\mu m$ bond lines is still under 1 ppm. In comparison to the analytical model using the linear approximation, the total power loss is also calculated analytically using Eq.~\ref{equ:final}. The percentage difference is shown on the right in Fig.~\ref{fig: PLXY_micro_compare}. We see that the linear approximation gives quite accurate results. For instance, in the displacement parameter region $(x_{0}, \theta)$ that gives 1 ppm power loss, the difference is roughly 2\%. The residual becomes large as we move toward larger displacements, as the linear approximation starts to fail.

\begin{figure}[htbp]
    \centering
    \includegraphics[width=0.9\linewidth]{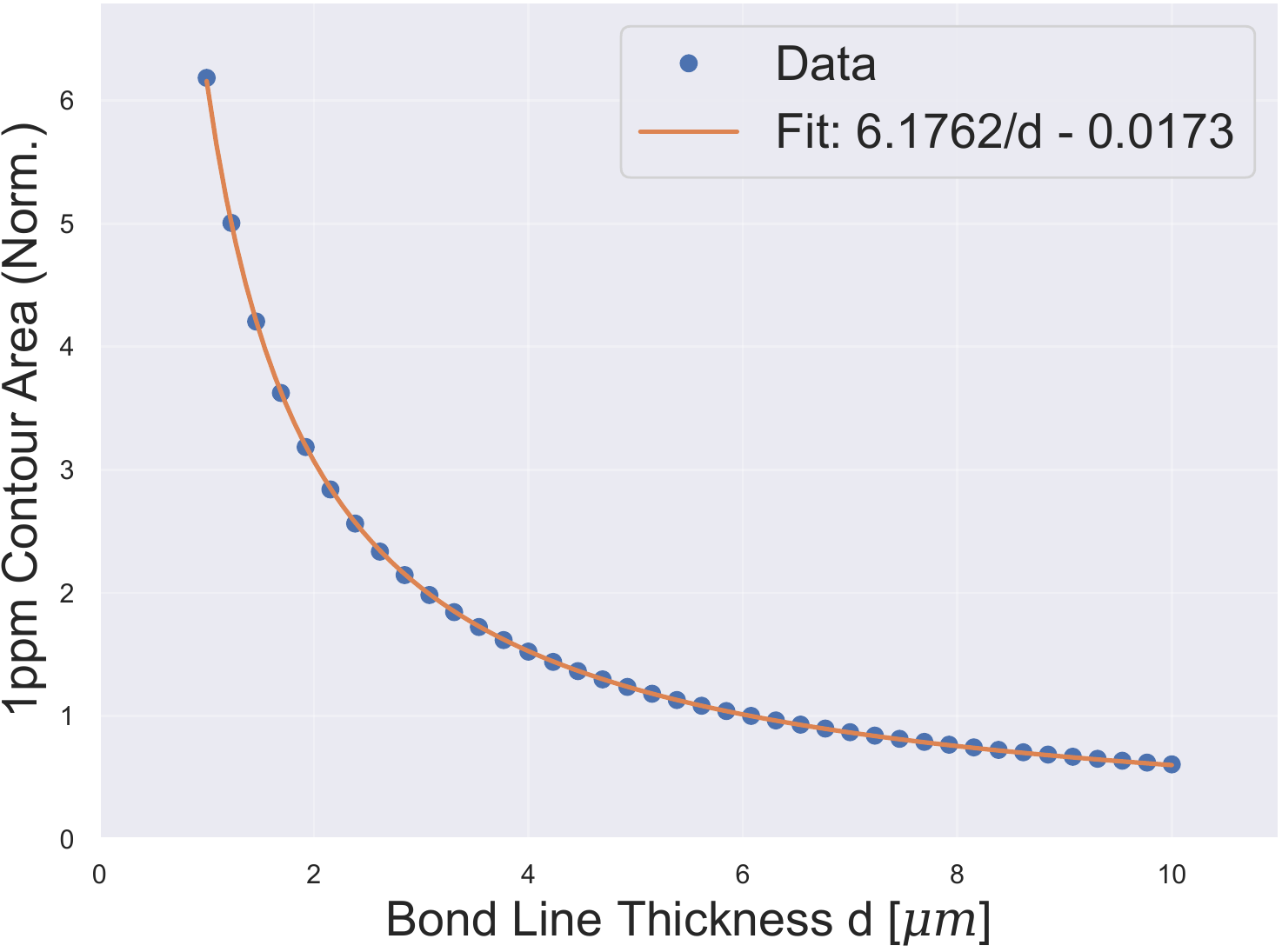}
    \caption{An image showing the 1 ppm contour area (normalized so that the value is 1 for the 6 $\mu m$ bond line case) in Fig.~\ref{fig: PLXY_micro_compare} as the bond line thickness is increased from 1 $\mu m$ to 10 $\mu m$. The best-fit function suggests a simple inverse relationship.}
    \label{fig: area_d}
\end{figure}

The beam displacement parameter contour tells us how much a beam can be displaced while keeping the total power loss within a small value. We quantify the beam displacement tolerance as the area of the contour. As explained earlier in Eq.~\ref{equ: bondlined}, the ``effective'' bond line thickness is uncertain, due to the uncertainty in the beam propagation direction. With the bond line thickness ranging from 1 $\mu m$ to 10 $\mu m$, the beam displacement contour area is calculated and shown in Fig.~\ref{fig: area_d}. The contour area is normalized such that the value is 1 for the 6 $\mu m$ bond line case we looked at. The best-fit function suggests a simple inverse relationship between the contour area and the bond line thickness, as shown in orange. For instance, if the bond line width is 3 $\mu m$ instead, we could tolerate twice as much beam displacement compared to the 6 $\mu m$ bond line case.

We can also explain the $1/d$ relation for the contour area using our analytical formalism. As we square and integrate the mode contents in Eq.~\ref{equ:final} to get the power loss, the terms that are linear in $x_{0}$ or $\theta$ cancel out due to the property of odd functions. For the quadratic terms, the coefficients in general depend on $d$. In our case, since $d$ is $\mathcal{O}( 1 \mu m)$ while $x_{0}$ is $\mathcal{O} (4\% \cdot 0.0394 m) \sim 1 mm$, we have $\mathrm{d} \ll x_{0}$. As a result, only the linear term in $d$ would contribute to the coefficients of the quadratic terms. Namely, we would have $C_{1} d \cdot x_{0}^{2} + C_{2} d \cdot \theta^{2} \sim 1 $, where $C_{1, 2}$ are constants. This is a function for an ellipse, as shown in Fig.~\ref{fig: PLXY_micro_compare}. The area equals to $\frac{\pi}{C_{1}C_{2} d}$, which is the inverse relation illustrated in Fig.~\ref{fig: area_d}.

This paper has demonstrated an assessment technique for evaluating the beam displacement tolerance performance benefit for a given optical mode and a segmented mirror geometry. It has shown that odd-indexed HG modes remain compatible with the segmented mirror idea, even when there is a substantial amount of beam displacement present. For a nominal bond line effective thickness of 6 $\mu m$, the total power loss for $\mathrm{HG}_{3,3}$ mode is within 1 ppm even with 1-degree rotation and 4\% its beam size lateral offset. There still remains a huge benefit compared to the fundamental $\mathrm{HG}_{0,0}$ mode, which has a total power loss of 170 ppm even when perfectly centered. 

These results have important implications for future ground-based gravitational wave detector designs relying on the use of high-purity silicon substrates for the test mass material~\cite{Adhikari_2020, ET}: they show that odd-indexed higher-order HG modes allow the use of segmented mirrors with overall diameter larger than the maximum available silicon boule diameter, by keeping the optical power on the bonds orders of magnitude smaller than for the $\mathrm{HG}_{0,0}$ mode.

\section*{Acknowledgments}
This work was supported by National Science Foundation grants PHY-1806461 and PHY-2012021.

\newpage
\bibliography{reference}

\end{document}